# Fast rotating Blue Stragglers prefer loose clusters

Francesco R. Ferraro[1,2*], Alessio Mucciarelli[1,2], Barbara Lanzoni[1,2], Cristina Pallanca[1,2], Mario Cadelano[1,2], Alex Billi[1,2], Alison Sills[3], Enrico Vesperini[4], Emanuele Dalessandro[2], Giacomo Beccari[5], Lorenzo Monaco[6], Mario Mateo[7]

[1]*Dipartimento di Fisica e Astronomia "Augusto Righi", Alma Mater Studiorum Universita` di Bologna, Via Piero Gobetti 93/2, I-40129 Bologna, Italy*

[2]*INAF -- Astrophysics and Space Science Observatory Bologna, Via Piero Gobetti 93/3, I-40129 Bologna, Italy*

[3]*Department of Physics & Astronomy, McMaster University, 1280 Main Street West, Hamilton ON, L8S 4M1, CANADA*

[4]*Department of Astronomy, Indiana University, Bloomington, IN, 47401, USA*

[5]*European Southern Observatory, Karl-Schwarzschild-Strasse 2, 85748 Garching bei Munchen, Germany*

[6]*Instituto de Astrofísica, Facultad de Ciencias Exactas, Universidad Andres Bello, Sede Concepcion, Talcahuano, Chile*

[7]*Department of Astronomy, University of Michigan, 1085 S. University, Ann Arbor, MI 48109,USA*

**Blue stragglers are anomalously luminous core hydrogen-burning stars formed through mass-transfer in binary/triple systems and stellar collisions. Their physical and evolutionary properties are largely unknown and unconstrained. Here we analyze 320 high-resolution spectra of blue stragglers collected in eight galactic globular clusters with different structural characteristics and show evidence that the fraction of fast rotating blue stragglers (with rotational velocities larger than 40 km/s) increases for decreasing central density of the host system. This trend suggests that fast spinning blue stragglers prefer low-density environments and promises to open an unexplored route towards understanding the evolutionary processes of these stars. Since large rotation rates are expected in the early stages of both formation channels, our results provide direct evidence for recent blue straggler formation activity in low-density environments and put strong constraints on the timescale of the collisional blue straggler slow-down processes.**

## Introduction

Blue straggler stars (BSSs) are puzzling objects well distinguishable in the colour-magnitude diagram (CMD) of stellar systems, where they define a sequence extending brighter and bluer than the Main sequence (MS) Turn-Off (TO) point, mimicking a sub-population of young (or more massive) stars[1-3]. Since no recent star formation has occurred in old globular clusters (GCs), mass-enhancement processes must be at the origin of BSSs and two main scenarios are currently favoured: stellar mergers induced by direct collisions[4-6], and mass-transfer activity in binaries[7-9], possibly triggered by stellar interactions. Because of this, BSSs are among the most massive objects populating star clusters: in old stellar systems as the Galactic GCs, they are significantly heavier ($M_{BSS}$=1.2-1.5 $M_\odot$[10,11]) than the average population ($<m>$ =0.3 $M_\odot$). This implies that they are subject to dynamical friction, which makes them sink to the bottom of the potential well. Hence, these stars are powerful probes[12-18] of the internal dynamical evolution of collisional stellar systems. Indeed, GCs with different levels of dynamical evolution can be ranked on the basis of the central concentration of their BSS population with respect to lower-mass (normal) stars (see section "BSSs and cluster dynamical age" in Methods). Despite such a huge potential as dynamical probes, many questions concerning the formation and evolution of BSSs are still unanswered, and the theoretical models aimed at describing these objects and their link with cluster dynamics remain largely unconstrained by observations. In fact, although BSSs have been routinely observed for 70 years now, only little information about their physical properties (like chemical abundances and rotational velocities) has been collected so far. To address this issue, in 2006 we started a spectroscopic survey of the BSS populations in a sample of Galactic GCs with different structural properties[19-24]. That large set of high-resolution spectra (see Supplementary Information) has led to the discovery of a sub-sample of BSSs with a depletion of carbon and oxygen in 47 Tucanae[19] and in M30[20], a feature that is considered a chemical signature of the mass-transfer formation channel.

Here we analyse the BSS rotational properties in 8 Galactic GCs: 47 Tucanae[19], M30[20], NGC 6397[21], M4[22], NGC 6752[23], ω Centauri[24], M55 and NGC 3201 (Supplementary Table 1 and 2 and Supplementary Fig. 1). We find that BSSs with rotational velocities larger than 40 km/s prefer low-density environments, suggesting recent formation of BSSs likely originated from mass-transfer activity in binary systems. Our results also provide constrains to the timescale of the collisional BSS slow-down processes.

## Results

DEFINING FAST SPINNING BLUE STRAGGLERS

Considering the entire dataset, the rotational velocities of a total of 320 BSSs have been measured in 8 GCs. Their distribution is plotted in the top panel of Figure 1a. To demonstrate the reliability of the rotational velocity measures, the spectra of 30-35 normal stars in each cluster have been secured with the same observational setups and analysed with the same procedures. The control sample is made of a few MSTO stars (about 10% of the total) and, mainly, red giant branch (RGB) and sub-giant branch (SGB) stars that are known to show negligible rotation. The distribution of their rotational velocities is plotted in the bottom panel of Figure 1a: as expected, all the measured values are smaller than 20 km/s. The difference between the distributions shown in the two panels is evident: 100% of the reference stars show negligible rotation, while only 54% of the surveyed BSSs have rotation velocities smaller than this value and the remaining 46% of the sample is distributed over a long tail extending toward much larger values, reaching (and, in a few cases even exceeding) 200 km/s. The BSS distribution decreases rather smoothly from 20 to 40-50 km/s, and a substantial portion of the population, approximately 28%, has rotation velocities larger than 40 km/s. This result confirms, once more, that BSSs are a peculiar sub-population, with properties that deviate from those observed in normal cluster stars.

To investigate whether this is the typical distribution of BSS rotation velocities, in Figure 1b we plot the results obtained in each surveyed cluster, separately. Although the relatively small number of BSSs observed in each system prevents statistically significant comparisons on a cluster-to-cluster basis, the figure shows clear differences in the fraction of fast rotating BSSs, and it reveals an unexpected, intriguing feature: two separated groups of GCs can be distinguished, one that harbours a modest fraction of fast rotating BSSs (namely, 47 Tucanae, M30, NGC 6397, and NGC 6752), and another group with a rotation velocity distribution more extended toward large values (including, M55, M4, ω Centauri, and NGC 3201), with a few objects[20-24] spinning even faster than 200 km/s (see also Supplementary Figure 2). Surprisingly, this cluster classification, based on the distribution of BSS rotational velocities, mirrors the grouping in terms of structural parameters: only low-density ($\log \rho_0 < 4$ in units of $L_\odot/pc^3$) and low concentration ($c<1.8$) systems show a relevant fraction of fast rotating BSSs. The effect is even more evident when the rotational velocity distributions of the two groups are directly compared (see top and bottom panels of Figure 2a). Undoubtedly, the long tail of the BSS rotational velocity distribution shown in Figure 1a is due to loose clusters, while almost the entire (96%) BSS population of high-density systems show rotation velocities smaller than 40 km/s. The comparison between the normalized cumulative distributions of rotation velocities shown in Figure 2b even more highlights the difference: while in high-density clusters only 4% (4 out of 92) of BSSs spin faster than 40 km/s, more than one third of the population (38%, i.e., 87 out of 228) rotates faster than this value in low-density environments. A Kolmorogov-Smirnov test quantifies the statistical significance of the difference: the probability that the two distributions are extracted from the same parent family is essentially zero (approximately $10^{-8}$), which corresponds to a significance level well above $5\sigma$. Thus, the main evidence emerging from this analysis is that the fraction of fast rotating BSSs appears to be larger in low-density, than in high-density GCs.

CORRELATIONS WITH CLUSTER PARAMETERS

The observed sample covers nearly the entire range of values expected for GCs in terms of structural parameters (for instance, the King concentration parameter c goes from 0.8 to 2.5, and the central luminosity density spans approximately 4 orders of magnitude, from $\log \rho_0 = 2$, to 6 in units of $L_\odot/pc^3$; see Supplementary Table1), thus providing an ideal dataset for exploring possible links between the chemical and rotational properties of BSSs, and the characteristics of their host clusters. To probe this connection, we computed the fraction of fast rotators (FRs) in each observed cluster and searched for correlations with other parameters. Although the exact definition of FR is somehow arbitrary, the distributions shown in Figures 1 and 2 suggest that a threshold value ranging from 30 to 50 km/s is appropriate. Thus, we define FRs as the BSSs with rotation velocity larger than (or equal to) 40 km/s. Figures 3a and 3b show the specific fraction of FRs (i.e., the number of FRs divided by the total number of surveyed BSSs: $f_{FR} = N_{FR}/N_{TOT}$) as a function of the King concentration parameter (c) and the central luminosity density ($\log \rho_0$) of the parent cluster. As can be seen, a strong correlation with the environment emerges from this study: the fraction of FRs steadily decreases from about 50%, to zero for increasing values of both c and $\log \rho_0$. The assumption of slightly different thresholds (30 or 50 km/s) has no impact on the results (see Supplementary Figure 3). Similar trends are detected with both the collisional parameter $\Gamma_{coll}$ (see Figure 3c), and the $A^+$ parameter (Figure 3d). The former provides a measure of the overall level of collisionality of the system (see Section "The collisional parameter" in Methods) and, by definition, increases with $\rho_0$ and the cluster core radius. The empirical parameter $A^+$ quantifies the level of BSS central segregation due to the dynamical ageing of the host stellar system (see Section "BSSs and the cluster dynamical age" in Methods), with larger values of $A^+$ corresponding to dynamically older clusters, where dynamical friction has been effective in concentrating heavier objects toward the centre of the system. Hence, the latter relation suggests that the fraction of fast spinning BSSs also depends on the internal dynamical evolution of the host cluster. Particularly intriguing is the behaviour of $f_{FR}$ versus the $A^+$ parameter. Indeed, for low values of $A^+$ (corresponding to dynamically younger clusters) the overall trend of $f_{FR}$ closely resembles that observed as a function of the other cluster parameters (namely, $f_{FR}$ decreases for increasing $A^+$; dashed line in the right-bottom panel). Then, the trend seems to change at $A^+ = 0.30$, above which $f_{FR}$ slightly increases with $A^+$ (dotted line in the same panel). According to the total sample investigated so far, all the clusters classified as post-core collapse systems have $A^+$ above this value (see section "BSSs and cluster dynamical age" in the Methods). We can therefore speculate that this (admittedly modest) increase of $f_{FR}$ could be due to an enhancement of recently-generated collisional BSSs (which are mainly expected to form in the cluster centre) at the time of core collapse and during the core oscillations that may characterize the post-core collapse phase.

Thus, low-density environments, which are characteristic of dynamically-young clusters and where stellar collisions are less probable, are the ideal habitat for fast spinning BSSs. In turn, the low collision rate suggests that in these environments the dominant BSS formation channel is mass-transfer in binary systems, and a correlation between the fraction of FRs and the cluster binary content should therefore be expected. Indeed, Figure 4a confirms that the percentage of FRs correlates with the overall binary fraction[25,26] of the parent cluster (note that ω Centauri is excluded because no reliable binary fraction estimate has been found in the literature). Moreover, it is worth noticing that, out of the 4 FRs detected in high-density clusters, the two observed in 47 Tucanae and in M30 are classified as contact eclipsing binaries[27,28], thus indicating their mass-transfer origin. These results point toward interesting scenarios never explored before.

**Discussion**

The observational findings discussed above show that fast spinning BSSs preferentially reside in low-density clusters, where they are likely originated from mass-transfer activity in binary systems, while in conditions of high-density, high collision rate and advanced stages of dynamical evolution, essentially only slowly rotating BSSs are found.

From the theoretical point of view, high rotational velocities are expected at birth for BSSs formed through both the proposed channels, because angular momentum transfer concurs with mass transfer in binary systems, and because collisional proto-BSSs rapidly contract conserving their angular momentum[5,6]. In later stages of evolution, braking mechanisms (such as magnetic braking, disk locking, and mass loss) are expected to intervene and slow down the stars, with timescales and efficiencies that are still poorly understood[29,6]. Standard magnetic braking seems to be at work and slow down mass-transfer BSSs in timescales shorter than 1 Gyr. This is suggested by the fact that the relation between rotation periods and ages observed for these objects in open clusters and the Galactic field[30] is well reproduced by the models used to describe the spin-down rate of single, low-mass MS stars[31] (see Section "Rotation in single stars" in Methods), which share the same structure in terms of convective envelopes and magnetic fields. In turn, this result indicates that the rotation rate can be used as gyro-chronometer to estimate the stellar age (measuring, in the case of BSSs, the time since the end of mass transfer). The situation is less clear in the case of collisional formation, since this braking mechanism is expected to not operate in collisional products, which develop no convective envelope. Smoothed particle hydrodynamics simulations[6] of stellar collisions showed that efficient angular momentum loss must occur in the early phases of evolution in order to keep these objects in the BSS region of the CMD, and the magnetic locking of the star to a disc was pointed out as a possible braking mechanism for collisional BSSs. However, no physically motivated constraints to the characteristic timescale of the process was derived, leaving room for other alternative mechanisms able to cause an efficient decrease of angular momentum in the forming BSS. Hence, fast rotation should be the signature of recently formed BSSs (either from mass-transfer in binary systems, or through collisions), with the time scales of the subsequent braking mechanism(s) remaining essentially unconstrained in the case of collisions.

These predictions can be used to interpret the observational evidence collected so far. The study presented here has revealed a dramatic discrepancy in the abundance of FRs in different environments: FRs are one order of magnitude less abundant (about 4%) in high-density environments, than in low-density clusters (about 38%). The fact that most of the observed FRs are found in low-density clusters, where binaries are more abundant (Figure 4a), probably is the manifestation of a generation of BSSs originated via mass-transfer activity during the secular evolution of primordial binary systems and formed less than 1 Gyr ago (according to the currently accepted braking mechanism timescale). The paucity of FRs in high-density clusters, instead, indicate that no relevant recent BSS formation occurred (neither from the mass-transfer, nor from the collisional channel), and/or very effective braking mechanisms are at work. This is a very important point, worth of further comments.

The evidence that the fraction of FRs originated by the mass-transfer channel is very different from that found in low-density environments is somehow surprising, since the mass-transfer channel is expected to be continuously active in all environments, including high-density clusters[32]. Figure 4b,c suggests that the explanation resides in the different fractions of binary systems characterizing the two environments. In fact, both the FR fraction and the binary frequency nicely correlate with the cluster central density, and the combination of the two best-fit relations shown in Figure 4b,c corresponds to the solid line reported in Figure 4a. This relation properly reproduces the observed trend between the fractions of FRs and binary systems, suggesting that low-density environments are naturally effective in preserving primordial binaries and then use them for forming BSSs. Conversely, in high-density systems the mass-transfer channel for BSS formation is limited by the less abundant populations of binaries, which is likely explainable by a destructive action of stellar interactions occurring in highly collisional environments. It is worth of noticing that high-luminosity clusters have been observed[33-35] to harbour a fraction of BSSs smaller than that measured in low-luminosity systems, and this has been interpreted as another manifestation of the fact that binaries have been effectively disrupted in the former, where stellar encounters are more frequent. However, as show in Supplementary Figure 4, the fraction of BSSs with high rotational speed appears to be independent of the global cluster luminosity, thus making even more relevant the strong correlation between $f_{FR}$ and density detected here.

While a paucity of recently formed mass-transfer BSSs in high-density clusters can be ascribed to a low binary fraction, a relevant number of collisional BSSs is expected to form in these systems by recurrent dynamical interactions: indeed, about 10-100 collisional BSSs are expected[36] to have formed during the last Gyr in the clusters with the highest collision rates. This naturally raises the question of why this collisional component is not observed as FRs.

The paucity of FRs in high-density clusters indicates that efficient braking mechanisms are at work also for collisional BSSs, in agreement with previous expectations[6]. Such unknown spin-down mechanism might be linked to the intrinsic nature and structure of these objects (formed though collisions instead of mass-transfer) or depend on the local density. For instance, by considering that the main source of angular momentum loss in collisional BSSs is stellar winds, a possibility is that the frequent stellar interactions occurring in high-density GCs contribute to increase mass loss. In any case, the data presented in this study can be used to set some constraints on the timescale of this braking mechanism, suggesting that it should be shorter than 1-2 Gyr. In fact, in the case of M30, the bluest BSS sequence (tracing a sort of straight line in the CMD) is thought to be populated by BSSs simultaneously generated through collisions during the core collapse event, approximately 1-2 Gyr ago[13,32]. Seven BSSs lying along this (blue, collisional) sequence have been spectroscopically investigated and, among them, only one star is a FR (with rotation velocity $v \sin i$=90 km/s). This star is classified as a contact binary, thus pointing out that it is not a collisional, but a mass-transfer BSS contaminating the blue sequence (in agreement with recent models[37]). The average value of $v \sin i$ of the other 6 collisional BSSs measured along the blue sequence (20 km/s) is however larger than the average rotation velocity ($<v \sin i> = 7$ km/s) of the further 7 BSSs investigated in M30, which populate the red sequence and are thought to be formed by mass-transfer activity in binary systems. According to the quoted scenario, these findings would suggest that the slow-down process of the collisional BSSs formed during core collapse is still ongoing, and it has been able to decrease the initial rotational velocity to the current value of approximately 20 km/s in a timescale of 1-2 Gyr. The example of M30 suggests that the line of investigation proposed in this paper could provide a new set of empirical constrains to the timescale of the collisional BSS slow-down processes. This methodology needs to be applied to other post-core collapse stellar systems, where a (blue) sequence of collisional BSSs has been detected (see the cases of NGC 362[18] and M15[38]).

This work has revealed a clear-cut connection between the fraction of fast rotating BSSs and the environment (see Figure 3). Hence, this effect should be visible even within an individual cluster.

Indeed, in GCs the local density dramatically changes (by orders of magnitude) from the centre to the periphery. Thus, at least in principle, this effect should be detectable. However, GCs are not the ideal objects where this test can be carried out because they are dynamically-alive stellar systems, where dynamical processes like dynamical friction and mass segregation favour the progressive migration of heavy objects toward the centre, therefore preventing to trace the native radial distribution of BSSs from observations. Moreover, such a study would require the spectroscopic screening and the measure of BSS rotation velocities along the entire radial extension of the cluster. Luckily, among the GCs investigated in this study, we included ω Centauri, which turns out to be the perfect stellar system where the test can be carried out. In fact, there is evidence that this stellar system is poorly dynamically evolved[12,15,39,40], thus indicating that dynamical processes have not had enough time to significantly alter the native radial distribution of massive stars (as BSSs). In addition, we collected 109 spectra of BSSs sampling a relevant portion of the cluster (up to approximately 13 arcmin, 5 $r_c$, with $r_c$=153"). By splitting the observed data set in two sub-samples (with r<2.5 $r_c$ and r>2.5$r_c$, respectively), we found that the FR fraction increases from 32% in the innermost region, to more than 50% in the periphery. A more fine-tuned investigation shows that the fraction of FRs reaches more than 60% at r>3$r_c$, where the stellar density is 1.5 orders of magnitude lower than in the centre. Figure 5 shows the normalized radial distributions of fast spinning BSSs (with v sin $i$≥ 40 km/s) and slowly rotating BSSs (with v sin $i$<40 km/s) in the cluster. The Kolmogorov-Smirnov test confirms that the two distributions are different at 2.3 σ level of significance. Thus, evidence that fast-rotating BSSs prefer low-density environments is found also within an individual stellar system.

The results presented in this paper identify low-density environments as the natural habitat for fast spinning BSSs. By taking into account that high rotational velocity is expected in the earliest phases of BSS evolution, the discovered relationships provide information about the current BSS formation activity in different clusters, suggesting that the formation of mass-transfer BSSs is fully ongoing in loose GCs (due to the secular evolution of primordial binaries), while it is significantly reduced in high-density environments, due to the lower fraction of survived binaries in these environments. Appropriate models of mass-transfer evolution are now required to obtain reliable age estimates for each of these stars from their position in the CMD. The combination of formation age and rotation velocity would provide the basic empirical quantities to identify the braking mechanism(s) needed to reproduce the observations and evaluate its/their efficiency (as shown in the case of open clusters[30]). For collisional BSSs, the observational evidence presented here call for an efficient braking mechanism acting on a timescale of 1-2 Gyr, thus soliciting for renewed observational and theoretical efforts to understand the physics of these stars. On the other side, additional observations are needed to better characterize the relation discovered here. In particular, only the two extreme ends of the distribution of GC central densities (i.e., log $\rho_0$<3.5 and log $\rho_0$>4.8 in units of $L_\odot/pc^3$) have been explored so far, and the question of what happens in between remains open. There could be either a sharp environmental density threshold below which BSSs can be FRs and above which they spin as normal cluster stars, or a smooth transition between the two regimes (like that drawn by the dashed line in Fig. 3b). Determining which kind of dependence exists between the local density and the BSS rotational velocities is the first mandatory step for the construction of models and interpretative scenarios able to provide a physical explanation of the puzzling evidence presented here. In turn, this would not only provide crucial information on the BSS physics (and possibly on the formation mechanisms of these exotic objects), but it would also shed new light on the effects of stellar interactions in collisional systems (like GCs), once more reinforcing the deep link between BSSs and the dynamical evolution of the host cluster[12-18].

# Methods

**BSSs and cluster dynamical age:** The level of BSS central segregation with respect to normal (lighter) stars has been quantified via the $A^+$ parameter[41,17], defined as the area between the cumulative radial distribution of BSSs and that of a lighter (reference) population, such as, e.g., red giant or horizontal branch stars. The strong correlation between this parameter and the number of central relaxation times experienced by the host cluster[15] demonstrates that $A^+$ provides a high-sensitivity empirical measure of the stage of dynamical evolution (the dynamical age) of the parent cluster, with low values of $A^+$ (0.1-0.2) indicating dynamically-young stellar systems, and large values of $A^+$ (larger than 0.2) corresponding to dynamically-evolved clusters. In particular, the analysis of the global sample (59 globular clusters) for which $A^+$ has been measured so far[15,42] shows that all the (8) explored post-core collapse clusters display an $A^+$ parameter larger than 0.30, thus suggesting the existence of a critical value ($A^+$=0.3) that flags the onset of core collapse.

**Data analysis:** The projected rotation velocities ($v \sin i$) have been measured in all the clusters by adopting the same technique. As a first step, for each target star the effective temperature and surface gravity have been estimated according to the position in the CMD. More specifically, a grid of isochrones[43] with ages varying between 100 Myr and several Gyr have been overplotted to the CMD, and the atmospheric parameters have been derived from the projection of each star onto the closest isochrone. Then, for each target a model atmosphere was generated with the code ATLAS9[44] and then a set of synthetic spectra with different values of rotation velocity has been simulated by using the code SYNTHE[45]. The synthetic spectrum that minimizes the $\chi 2$ with respect to the observed spectrum finally provides the rotation velocity of the target star.

The uncertainties on the measured rotation velocities have been estimated by performing Monte Carlo simulations for a representative grid of simulated spectra computed for different values of the rotation velocity and the atmospheric parameters, and by adding different levels of Poisson noise to also account for varying spectral quality (signal-to-noise ratio). For each combination of parameters, a sample of several hundred synthetic spectra were simulated, and the analysis repeated on each of them. The recovered rotation velocities trace a nearly Gaussian distribution, and we assumed as error on *v* sin *i* the sigma of the derived distribution. Typical uncertainties of the order of a few km/s have been obtained for slowly rotating stars, while the errors reach 10-15 km/s in the case of the highest FRs (with *v* sin *i* >100 km/s).

**The collisional parameter** – The encounter rate between single stars, in a cluster with central mass density $\rho_0$, core radius $r_c$ and velocity dispersion $\sigma$ can be expressed[46] as:

$$\Gamma_{coll} \propto \rho_0^2 \, r_c^3/\sigma \qquad (1)$$

The link between velocity dispersion, core mass and radius that holds for virialized systems following the King model[47] allows a simplified expression of the collisional parameter:

$$\Gamma_{coll} \propto \rho_0^{1.5} \, r_c^2 \qquad (2)$$

The values listed in Supplementary Table 1 and plotted in Figure 3c have been computed through Equation (2) assuming the values of core radius from previous papers[12,13,15] and converting them in physical sizes by adopting the cluster distances[48,49]. The constant of proportionality has been neglected and the quoted values of $\Gamma_{coll}$ therefore have no meaning in absolute terms, but they provide the ranking of the 8 clusters under investigation in terms of their overall level of collisionality. It is noteworthy that a precise estimate of the real collision rate in a star cluster should also take into account the binary-binary and single-binary events, which, depending on the binary fraction, can be dominant[36,50,51] because of the larger cross-section. However, the systematic decrease of the binary fraction for increasing central density observed in globular clusters (see Figure 4b) suggests that $\Gamma_{coll}$ reasonably well represents the real collision rate in high-density systems, while it could be a lower-limit in low-density environments (thus, the anti-correlation between $f_{FR}$ and $\Gamma_{coll}$ could be even steeper than that shown in Figure 3c).

**Rotation in single stars** - Recent surveys of rotational velocities of solar type stars (mainly with Kepler[52]) have demonstrated that, independently of their initial rotation rate at birth, after less than 1 Gyr stars with the same age converge to the same rotation rate, regardless of their initial angular momentum.

**Correspondence to:** F.R.Ferraro[1] Correspondence and requests for materials should be addressed to F.R.F. <u>francesco.ferraro3@unibo.it</u>

**Acknowledgements:** This research is part of the project COSMIC-LAB at the Physics and Astronomy Department of the Bologna University (see the web page: http://www.cosmic-lab.eu/Cosmic-Lab/Home.html ). The research has been funded by project *Light-on-Dark*, granted by the Italian MIUR through contract PRIN-2017K7REXT (FRF). The research is based on data acquired at the Very Large Telescope of the European Southern Observatory, Cerro Paranal (Chile), under proposals 072.D-0337; 073.D-0093; 077.D-0696; 081.D-0356; 087.D-0748; 089.D-0298; 089.D-0306; 093.D-0270. Part of the data have been collected at the Magellan Telescope at Las Campanas Observatory in Chile under proposal CN2019A-15.

**Author contributions:** FRF designed the study and coordinated the activity. AM coordinated the reduction of the entire database and CP, MC, AB, ED and GB contributed to the data analysis. LM and AM prepared, and MM executed the observations of NGC 3201. EV and AS were in charge of the theoretical model development. FRF and BL wrote the first draft of the paper. All the authors contributed to the discussion of the results and commented on the manuscript.

**Competing Interests:** The authors declare no competing interests.

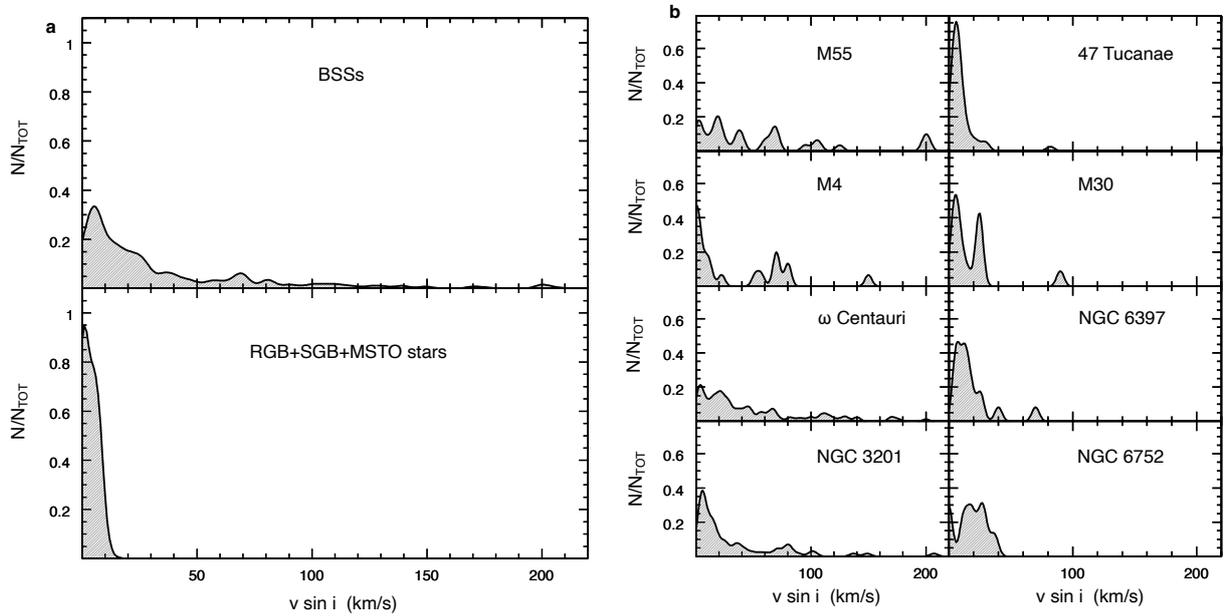

**Figure 1 – Comparing the distributions of BSS rotational velocities. a,** Kernel-density distributions of the rotational velocities measured for 320 BSSs in the 8 surveyed GCs (top panel), compared with that obtained for a sample of 216 RGB, SGB and MSTO stars (bottom panel). Both distributions are normalized to the total number of stars observed in each sample. **b,** Kernel-density distributions of the BSS rotational velocities observed in each of the eight clusters, normalized to the total number of BSSs observed in every system.

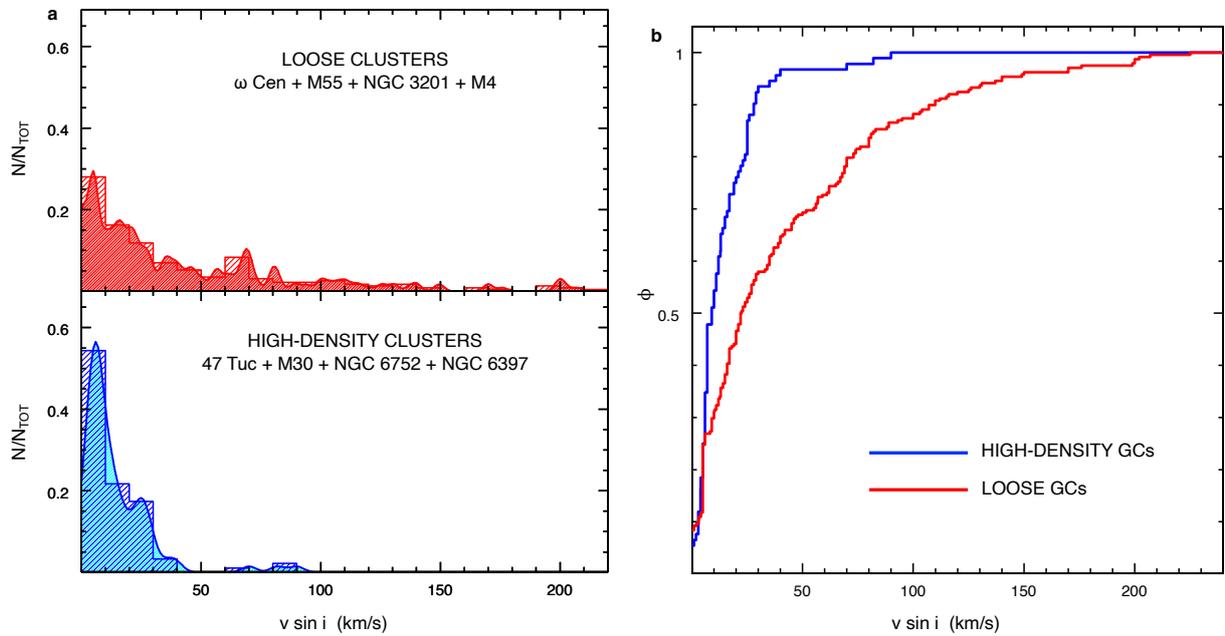

**Figure 2 – Comparing the distributions of BSS rotational velocities in loose and high-density clusters. a,** BSS rotational velocity distribution for loose clusters (namely, ω Centauri, M55, NGC3201 and M4), compared to that of high-density clusters (namely, 47 Tucanae, M30, NGC 6752 and NGC 6397), in the top panel (red histogram) and in the bottom panel (blue histogram), respectively. In both cases, the fraction is referred to the total number of BSSs observed in each sample. **b**, Comparison between the normalized cumulative distributions of BSS rotational velocities in low-density (red line) and high-density (blue line) clusters. A Kolmogorov-Smirnov test applied to the two distributions confirms that they are different at more than 5 σ significance level.

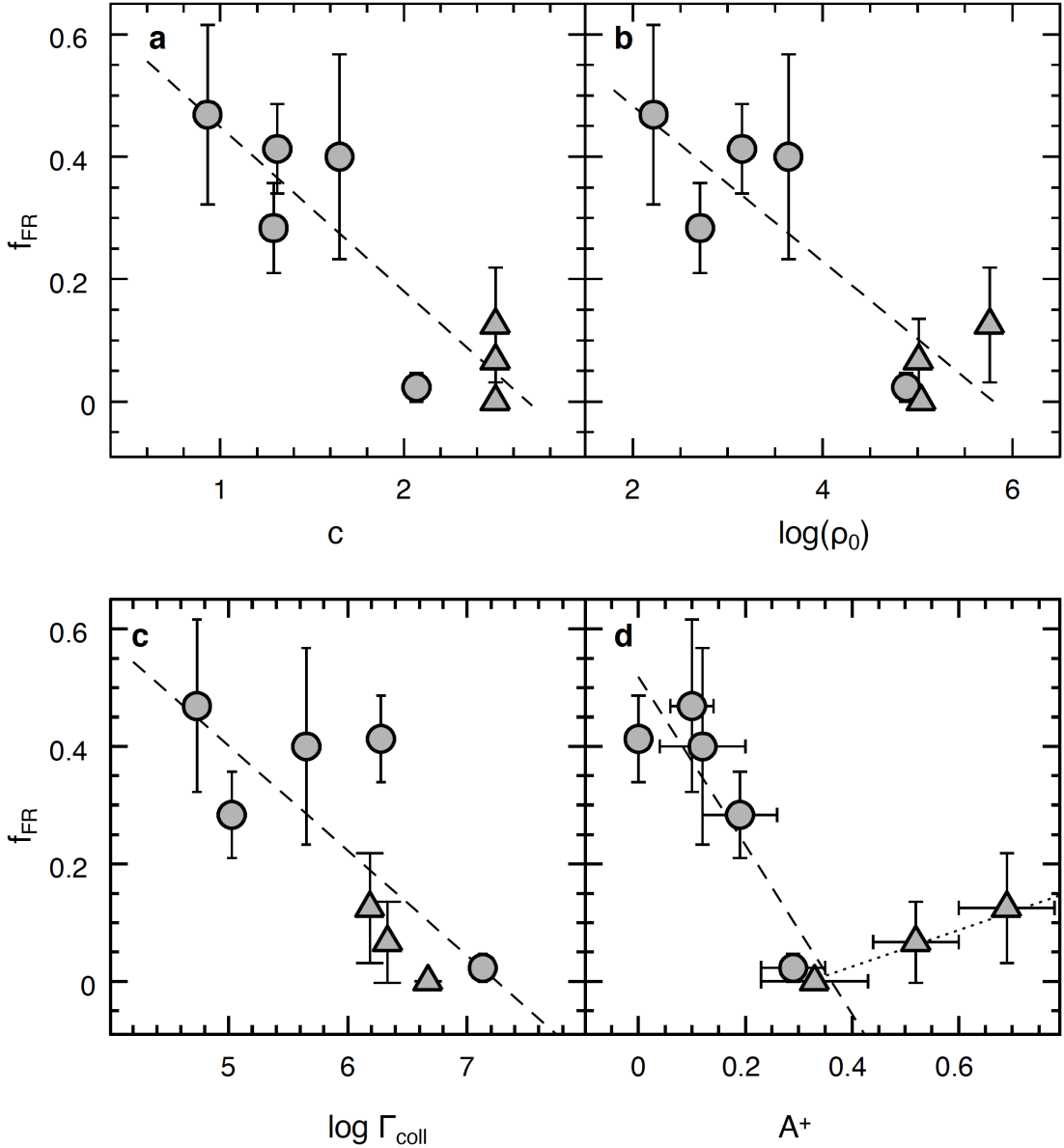

**Figure 3 – Relations between the observed fraction of fast rotators and the parent cluster characteristics.** Fraction of FRs, $f_{FR}$ (i.e., BSSs with $v \sin i \geq 40$ km/s; grey symbols) plotted as a function of the King concentration parameter ($c=\log(r_t/r_c)$, where $r_t$ is the tidal radius and $r_c$ is the core radius of the cluster, panel a), the central luminosity density ($\log \rho_0$ in units of $L_\odot/pc^3$; panel b), the collisional parameter ($\Gamma_{coll}$ in arbitrary units; panel c), and the cluster dynamical age as measured by the $A^+$ parameter (arbitrary units) from the central segregation of BSSs (panel d). The three clusters classified as core collapse systems (namely, M30, NGC 6397 and NGC 6752) are plotted as triangles. The errors computed following the Poisson statistics are also reported. The dashed lines are the best linear fits to the data (for $A^+ < 0.3$ in panel d). The values of the fit parameters and uncertainties are provided in the Supplementary Table 3. The dotted line drawn in the bottom-right panel illustrates the possible change of the $f_{FR} - A^+$ trend in the post core collapse state.

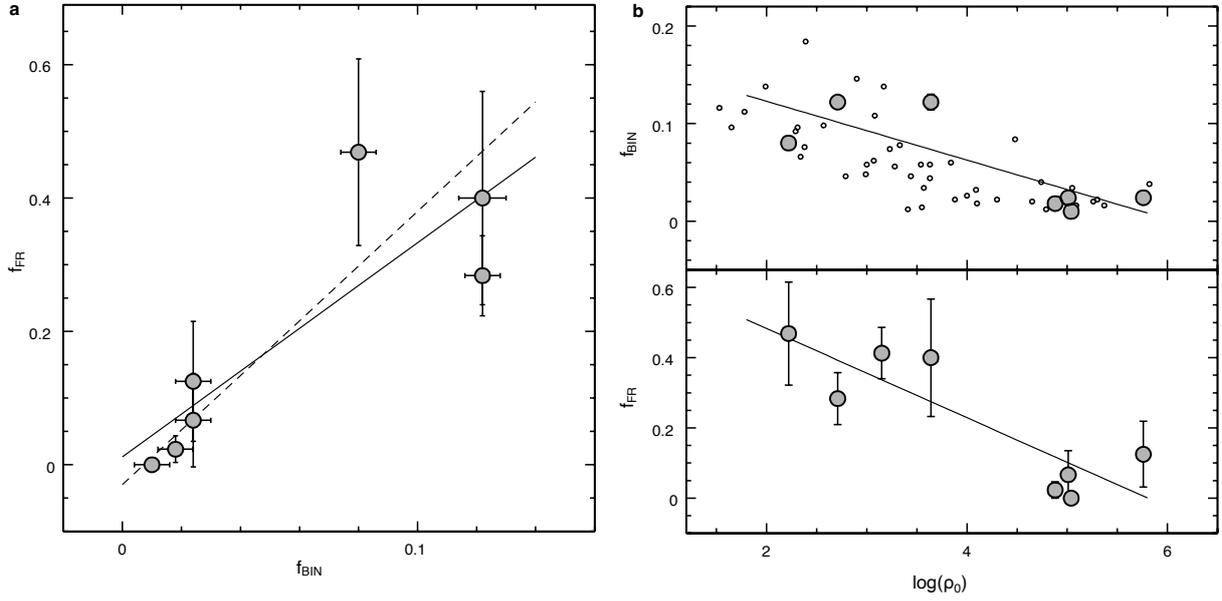

**Figure 4 –   The natural correlation between the fraction of FRs and the binary fraction. a,** Fraction of fast spinning BSSs, $f_{FR}$ (with $v \sin i \geq 40$ km/s; grey circles) plotted as a function of the global binary fraction[25] , $f_{BIN}$ (ω Centauri is lacking because no reliable estimates of its binary fraction are available in the literature).  Errors computed following the Poisson statistics are also reported. The dashed line is the best fit to the data, while the solid line traces the equation obtained by combining the two best fit relations shown in panel b. **b,** The relations linking the fraction of binaries $f_{BIN}$ (grey circles in the top panel) and the fraction of fast rotating BSSs $f_{FR}$ (grey circles in the bottom panel) to the cluster central density (log $\rho_0$ in units of $L_\odot/pc^3$). The small empty circles in the top panel are the data for a sample of 59 Galactic GCs[25]. The solid lines are the best linear fits to the data (see Supplementary Table 3 for the values of the fit parameters and uncertainties).

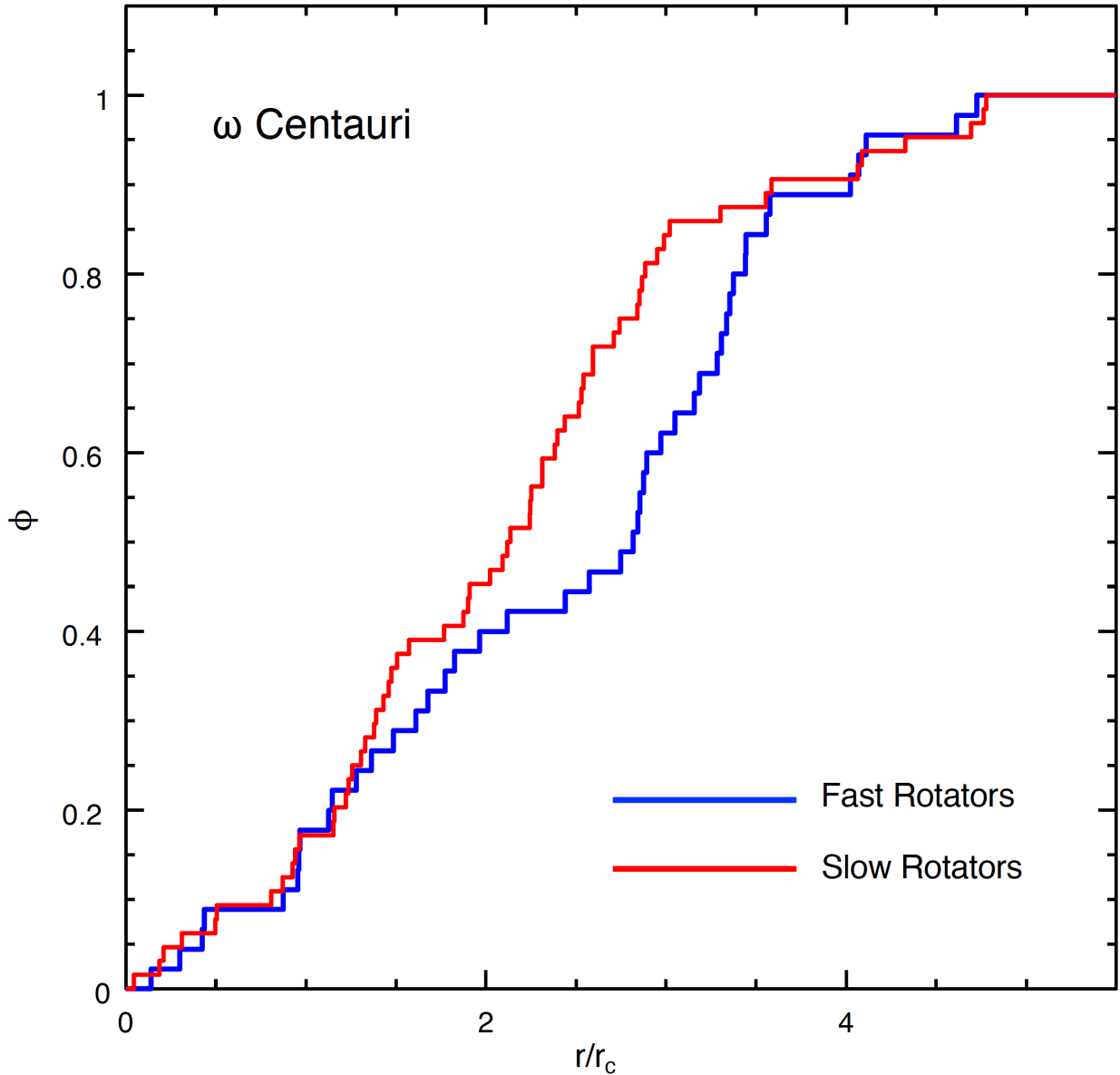

**Figure 5 – Confirming the trend within an individual cluster: the case of ω Centauri –** The normalized cumulative radial distributions (ϕ, arbitrary units) of the fast (blue line) and slowly (red line) rotating BSSs in ω Centauri as a function of the distance from the cluster center (r) expressed in units of the cluster core radius ($r_c$=153")[12]. The p-value for the Kolmorogov-Smirnov test is p=0.02, indicating a statistically significant difference between the two distributions at approximately 2.3 σ level.

# Supplementary Information for
# "Fast rotating Blue Stragglers prefer loose clusters"

Francesco R. Ferraro[1,2*], Alessio Mucciarelli[1,2], Barbara Lanzoni[1,2], Cristina Pallanca[1,2], Mario Cadelano[1,2], Alex Billi[1,2], Alison Sills[3], Enrico Vesperini[4], Emanuele Dalessandro[2], Giacomo Beccari[5], Lorenzo Monaco[6], Mario Mateo[7]

[1]Dipartimento di Fisica e Astronomia "Augusto Righi", Alma Mater Studiorum Universita` di Bologna, Via Piero Gobetti 93/2, I-40129 Bologna, Italy
[2]INAF -- Astrophysics and Space Science Observatory Bologna, Via Piero Gobetti 93/3, I-40129 Bologna, Italy
[3]Department of Physics & Astronomy, McMaster University, 1280 Main Street West, Hamilton ON, L8S 4M1, CANADA
[4]Department of Astronomy, Indiana University, Bloomington, IN, 47401, USA
[5]European Southern Observatory, Karl-Schwarzschild-Strasse 2, 85748 Garching bei Munchen, Germany
[6]Departamento de Ciencias Fisicas, Universidad Andres Bello, Fernandez Concha 700, Las Condes, Santiago, Chile
[7]Department of Astronomy, University of Michigan, 1085 S. University, Ann Arbor, MI 48109,USA

**The Data-set:** For this study we have used a collection of high-resolution spectra acquired over the last 15 years (see Supplementary Table 2). Most of the data have been acquired with the multi-object spectrograph FLAMES-GIRAFFE[1] mounted at the Very Large Telescope (VLT) of the European Southern Observatory (ESO). The instrument allows the simultaneous acquisition of high resolution (R=10000-30000) spectra for about 130 targets over a field of view of 25' in diameter. Part of the dataset has been already published and we refer to the original papers for the detailed description of the observations and data reduction: 47 Tucanae[2], M4[3], NGC 6397[4], M30[5], NGC 6752[6], ω Centauri[7]. For M55 the datasets have been acquired with FLAMES@VLT under program 093.D-0270 (PI: Lovisi), in June-July 2014, through 16 repeated exposures (each one of ~ 2760 sec) in the spectral region 3854-4049 Å, with a spectral resolution of 19,600.

In the case of NGC 3201 we used the multi-object fiber system Michigan/Magellan Fiber System (M2FS)[8], which feeds the double spectrograph MSPec mounted at the Magellan Clay Telescope at the Las Campanas Observatory, in Chile. This instrument allows the simultaneous observation of up to 128 objects per spectrograph over a field of view of about 30' of diameter. The data have been acquired in February-March 2019, through 16 repeated exposures in the spectral region 5127-5184 Å, with a spectral resolution of 18000, thus allowing us to properly sample the Mg lines at 5167.3 Å and 5172.6 Å.

**The spectroscopic sample-** The spectroscopic targets in each cluster have been selected to provide a reasonable sampling of the BSS population in both luminosity and radial distribution. The selection is based on extensive photometric catalogues (in most cases covering the entire cluster extension) published by our group over the years for each investigated system: ωCentauri[9], M30[10], 47Tucanae[11], NGC6752[12], M55[13], M4[3]. Typically, we used ultraviolet Hubble Space Telescope data in the central region (thus minimizing crowding issues) and ground-based optical observations for the outskirts. Of course, to keep a reasonably high signal-to-noise level, the faint extension of the BSS sequence is in general not included in the sample. In addition, to avoid possible biases due to effects of internal dynamical evolution, special care was devoted to properly sample the entire BSS radial distribution in each cluster, including the innermost regions where mass segregation processes and stellar interactions are particularly efficient. To avoid contamination of the spectra from scattered light, only isolated BSSs have been selected: all the BSSs having stars of comparable or brighter luminosity within a distance of about 3" (which is more than twice the fiber size) have been excluded from the selection. With these constraints, we succeed in safely allocating some fiber even in the central regions of high-density clusters. For instance, half of the spectroscopic sample (7 BSSs out of 15) has been observed within the innermost 15" from the centre of M30, which is a very concentrated, post-core collapse system. The number of surveyed BSSs is provided in Supplementary Table 1 and their distribution on the plane of the sky is plotted in Supplementary Figure 1, which clearly shows that, in all cases, the observed samples probe the central part of the cluster and extend to the external regions (out to 1-2 half-mass radii).

**Supplementary Table. 1 | The BSS sample and the main parameters of the target clusters.**

| Name | $N_{TOT}$ | $N_{FR}$ | [Fe/H] | $M_V$ | c | log $\rho_0$ | $\Gamma_{coll}$ | $A^+$ | Reference |
|---|---|---|---|---|---|---|---|---|---|
| M55 | 32 | 15 | -1.9 | -7.57 | 0.93 | 2.22 | $5.4 \times 10^4$ | 0.10 | This paper |
| NGC3201 | 67 | 19 | -1.6 | -7.45 | 1.29 | 2.71 | $1.0 \times 10^5$ | 0.19 | This paper |
| ω Centauri | 109 | 45 | -1.6 | -10.26 | 1.31 | 3.15 | $1.9 \times 10^6$ | 0.00 | Mucciarelli et al. (2013) |
| M4 | 20 | 8 | -1.2 | -7.19 | 1.65 | 3.64 | $4.5 \times 10^5$ | 0.12 | Lovisi et al. (2010) |
| 47 Tucanae | 43 | 1 | -0.7 | -9.42 | 2.1 | 4.88 | $1.4 \times 10^7$ | 0.29 | Ferraro et al. 2006 |
| M30 | 15 | 1 | -2.2 | -7.45 | 2.5 | 5.01 | $2.2 \times 10^6$ | 0.52 | Lovisi et al. (2013a) |
| NGC6752 | 18 | 0 | -1.5 | -7.73 | 2.5 | 5.04 | $4.7 \times 10^6$ | 0.33 | Lovisi et al. (2013b) |
| NGC6397 | 16 | 2 | -2.0 | -6.64 | 2.5 | 5.76 | $1.5 \times 10^6$ | 0.69 | Lovisi et al. (2012) |

The numbers of observed BSSs ($N_{TOT}$) and fast rotators ($N_{FR}$) are listed in columns 2 and 3, respectively, and the corresponding reference publication is reported in column 9. The values of iron abundance ([Fe/H]), integrated absolute magnitude ($M_V$), King concentration parameter (c), and central luminosity density (log $\rho_0$) are taken from the Harris Globular Cluster Catalog[14], those of the collisional parameter $\Gamma_{coll}$ have been computed as described in Methods, and those of the $A^+$ parameter are from the literature[15].

**Supplementary Table 2 | The observational data base**

| CLUSTER | PROGRAMME | PI | INTRUMENT | ONLINE DATA AVAILABILITY |
|---|---|---|---|---|
| 47 Tucanae | 072.D-0337 | FERRARO | ESO-VLT-FLAMES | 1 |
| ω Centauri | 077.D-0696 | FREYHAMMER | ESO-VLT- FLAMES | 1 |
| | 081.D-0356 | FERRARO | ESO-VLT- FLAMES | |
| | 089.D-0298 | FERRARO | ESO-VLT- FLAMES | |
| M4 | 081.D-0356 | FERRARO | ESO-VLT- FLAMES | 1 |
| M30 | 087.D-0748 | LOVISI | ESO-VLT-FLAMES | 1 |
| | 089.D-0306 | FERRARO | ESO-VLT-X-SHOOTER | |
| NGC 6397 | 073.D-0093 | FERRARO | ESO-VLT- FLAMES | 1 |
| | 081.D-0356 | FERRARO | ESO-VLT- FLAMES | |
| NGC 6752 | 081.D-0356 | FERRARO | ESO-VLT- FLAMES | 1 |
| | 089.D-0298 | FERRARO | ESO-VLT- FLAMES | |
| M55 | 093.D-0270 | LOVISI | ESO-VLT- FLAMES | 1 |
| NGC 3201 | CN2019A-15 | MONACO | MAGELLAN-M2FS | 2 |

1. http://archive.eso.org/eso/eso_archive_main.html
2. http://www.cosmic-lab.eu/Cosmic-Lab/BSS_rotation_spectra.html

**Supplementary Table 3 | The main best-fit relations.** Parameters and uncertainties of the best-fit relations shown as solid lines in the figures listed in the first column.

| Figure | Best fit relations | rms |
|---|---|---|
| Figure 3a | $f_{FR} = -0.26 (\pm 0.06) \times c + 0.72 (\pm 0.12)$ | 0.09 |
| Figure 3b, 4b | $f_{FR} = -0.12 (\pm 0.03) \times \log(\rho_0) + 0.74 (\pm 0.14)$ | 0.10 |
| Figure 3c | $f_{FR} = -0.18 (\pm 0.06) \times \log(\Gamma_{coll}) + 1.29 (\pm 0.39)$ | 0.12 |
| Figure 3d | $f_{FR} = -1.43 (\pm 0.46) \times A^+ + 0.52 (\pm 0.08)$ | 0.08 |
| Figure 4a | $f_{FR} = 3.21 (\pm 0.97) \times f_{BIN} + 0.01 (\pm 0.07)$ | 0.10 |
| Figure 4b | $f_{BIN} = -0.03 (\pm 0.01) \times \log(\rho_0) + 0.18 (\pm 0.04)$ | 0.03 |

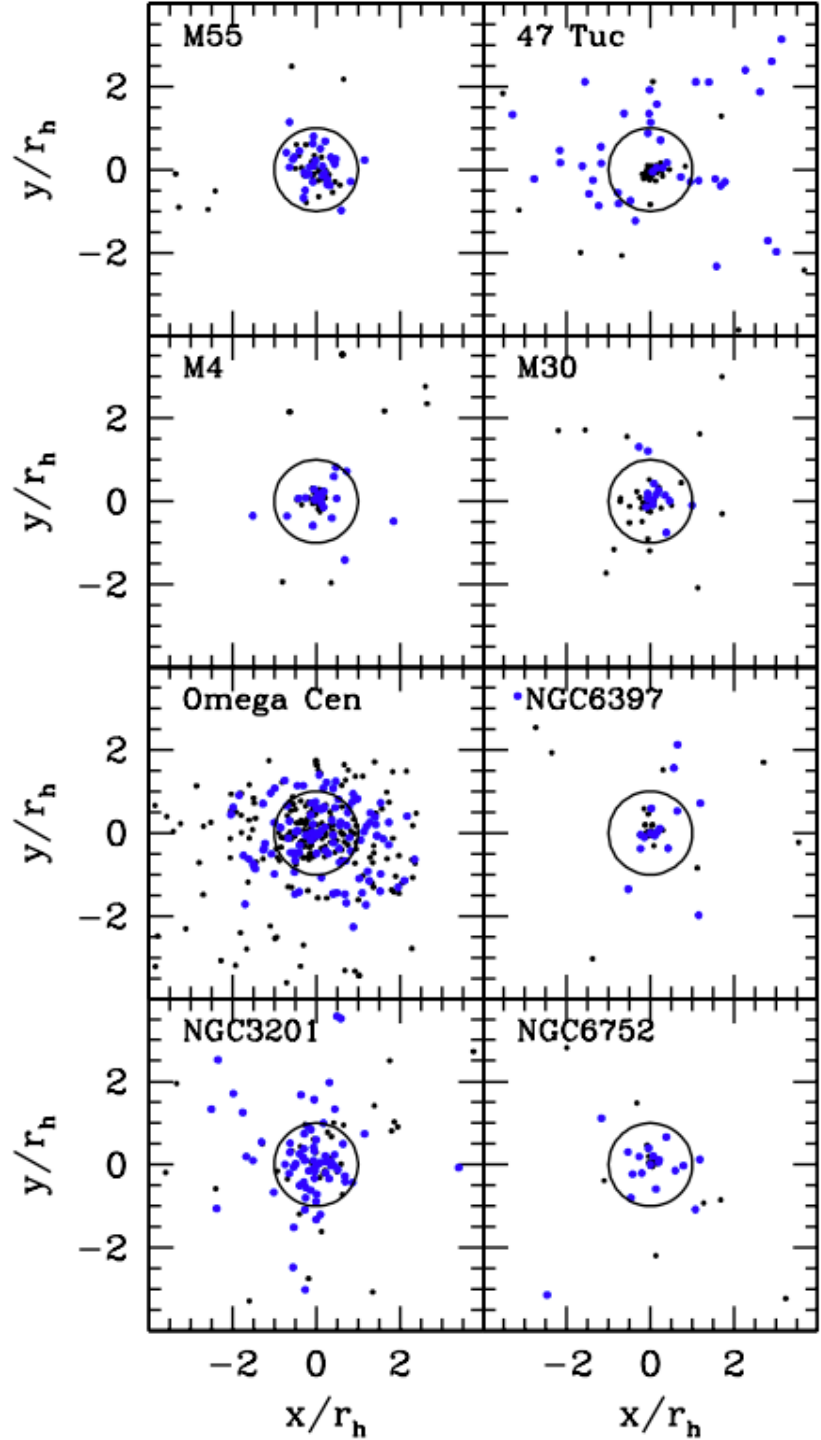

**Supplementary Fig. 1 – The radial distribution of the spectroscopic sample.** The sky distribution of the BSSs with measured rotation velocity (blue circles) is shown for each target cluster. The BSSs not observed in the spectroscopic survey are shown as black dots and are not included in the statistics. The x and y distances from the centre are plotted in units of the cluster half-mass radius ($r_h$). The black circles have radius equal to one $r_h$.

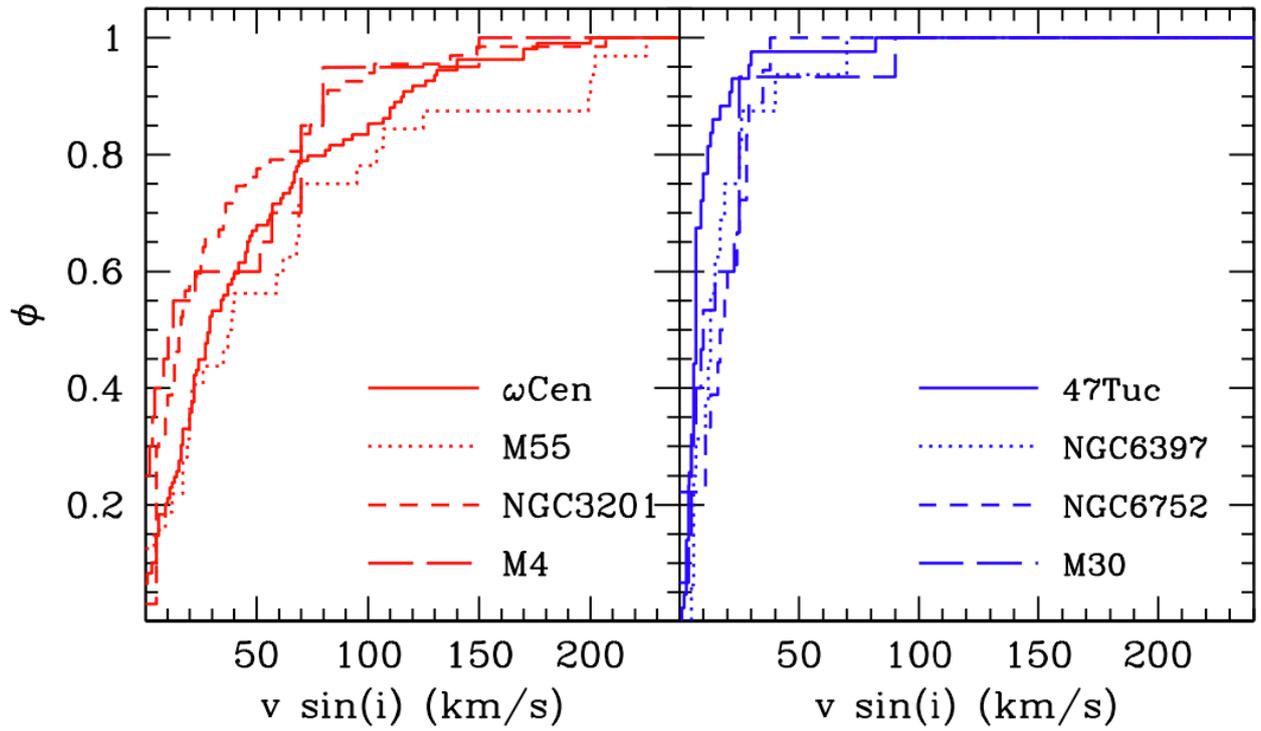

**Supplementary Fig. 2 – Comparing the rotation velocity distributions.** The normalized cumulative distributions ( ϕ ) of rotation velocities (v sin i) measured in the programme clusters. The four low-density clusters in the left panel (namely ωCentauri, M55, NGC 3201 and M4) show shallower cumulative distributions (indicating larger numbers of fast rotating BSSs), than the four high-density clusters in the right panel (namely 47 Tucanae, NGC 6397, NGC 6752 and M30).

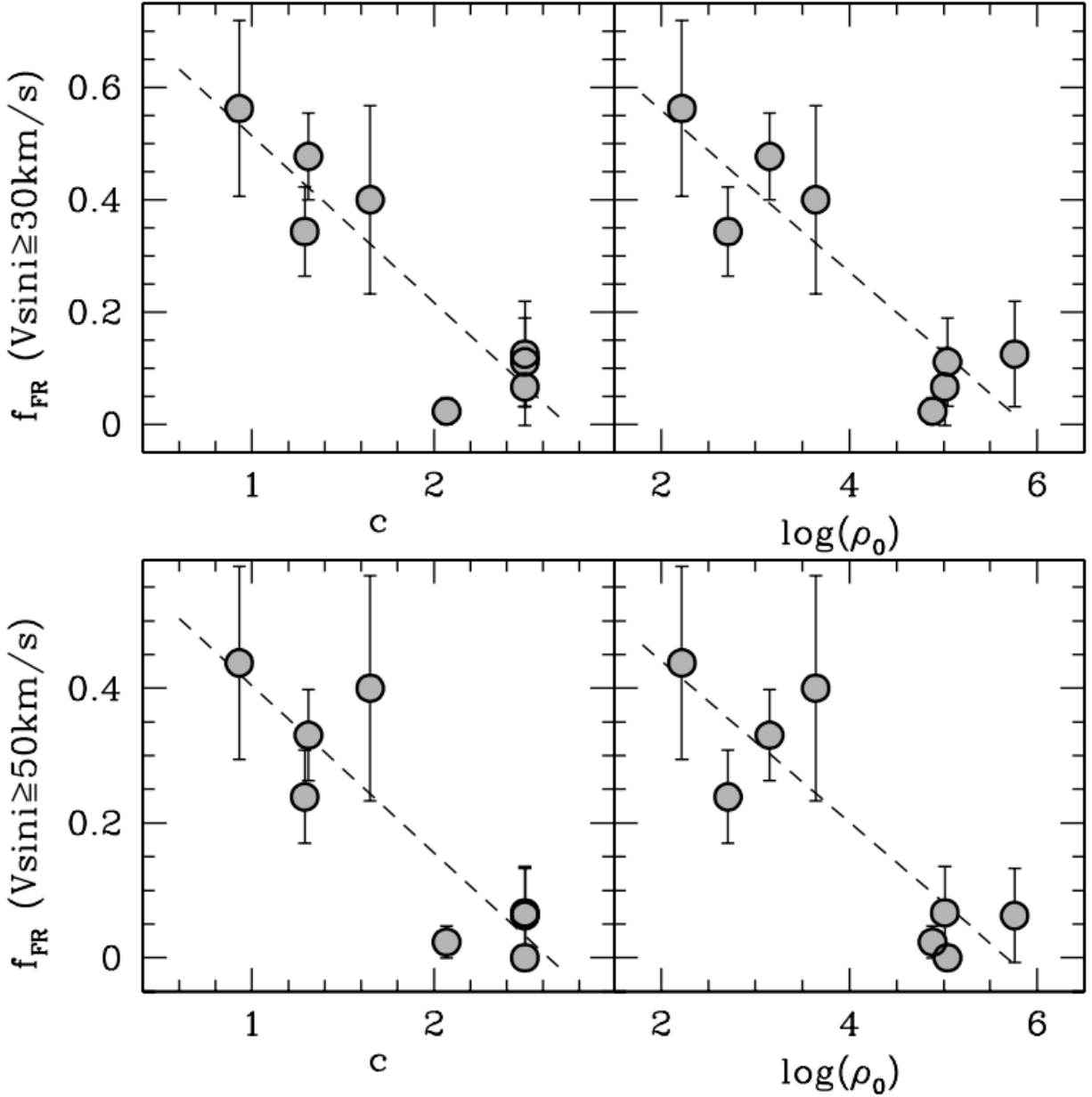

**Supplementary Fig. 3 – Changing the FR threshold.** Fraction of FRs, $f_{FR}$ (grey circles) as a function of the cluster concentration ($c=\log(r_t/r_c)$, where $r_t$ is the tidal radius and $r_c$ is the core radius of the cluster, left panels) and central luminosity density ($\log \rho_0$ in units of $L_\odot/pc^3$; right panels) for two different assumptions of the threshold used to define FRs: 50 km/s (top panels) and 30 km/s (bottom panels). The comparison between this figure and Figure 3 clearly shows that the overall characteristics of the distribution remain unchanged independently of the adopted threshold. The errors are computed following the Poisson statistics. The dashed lines are the linear best fits to the data.

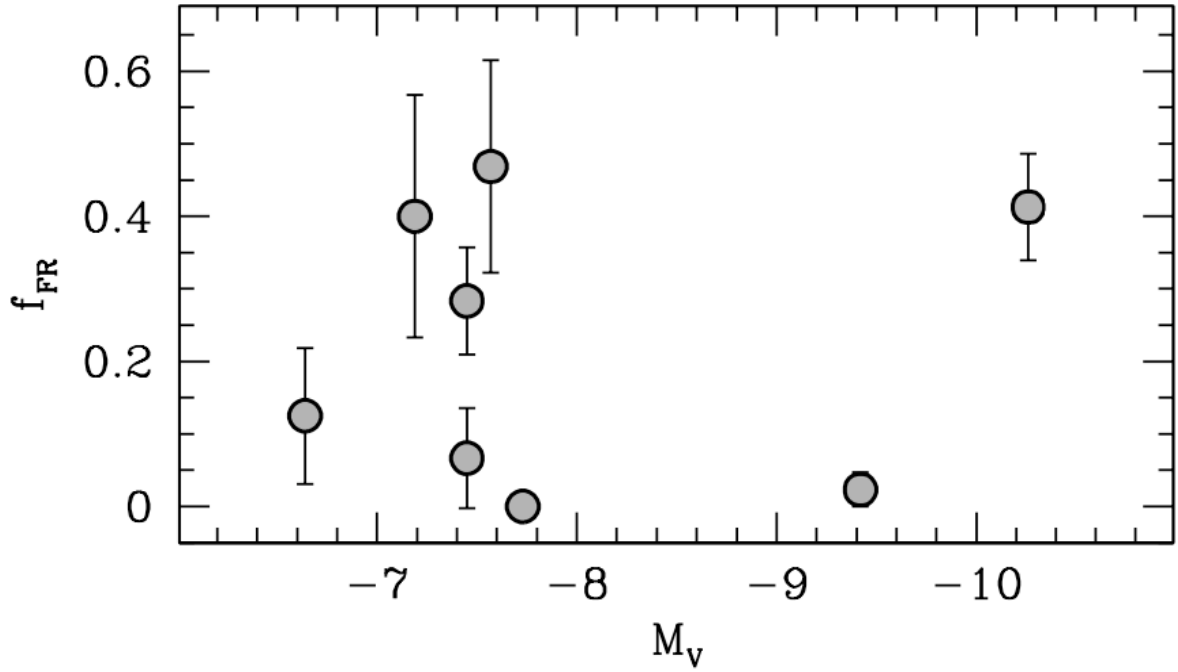

**Supplementary Fig. 4 – An orthogonal information.** While the overall fraction of BSSs is found[16-18] to decrease for increasing cluster luminosity, the fraction of FRs shows no significant correlation with this property ($M_V$ is the integrated absolute magnitude of the surveyed clusters). The errors are computed following the Poisson statistics.